# Band bending and ratcheting explain triboelectricity in a flexoelectric contact diode

K. P. Olson, C. A. Mizzi, L. D. Marks[*]

Department of Materials Science and Engineering, Northwestern University; Evanston, Illinois, United States of America.

*Corresponding author. Email: l-marks@northwestern.edu

**Abstract**

Triboelectricity was recognized millennia ago, but the fundamental mechanism of charge transfer is still not understood. We have recently proposed a model where flexoelectric band bending due to local asperity contacts drives triboelectric charge transfer in non-metals. While this ab-initio model is consistent with a wide range of observed phenomena, to date there have been no quantitative analyses of the proposed band bending. In this work we use a $Pt_{0.8}Ir_{0.2}$ conductive atomic force microscope probe to simultaneously deform a Nb-doped $SrTiO_3$ sample and collect current-bias data. The current that one expects based upon an analysis including the relevant flexoelectric band-bending for a deformed semiconductor quantitively agrees with the experiments. The analysis indicates a general ratcheting mechanism for triboelectric transfer and strong experimental evidence that flexoelectric band-bending is of fundamental importance for triboelectric contacts.

Triboelectricity is a wide-spread phenomenon of importance in areas ranging from the processing of pharmaceutical powders[1-3], nanoscale energy harvesting[4-9], the electrification of blowing sand, snow or volcanic plumes[10-14], damage to wind turbines[15], combing human hair[16,17] and even planetary formation[18]. It occurs whenever two materials rub together or discrete particles collide – the latter is often called contact or granular electrification.

The current literature includes older experiments where hypotheses such as the "triboelectric series" were shown to be false[19-24], to recent work in specialized areas which sometimes miss the earlier work. It is generally accepted that the triboelectric effect depends upon transfer of electrons[25-27], ions[28,29], and/or charged molecular fragments[30]. While there is some experimental evidence in support of many of these processes, at present there is little consensus in the literature. For instance, one common idea is that differences in the work function drive charge transfer, often called the Volta-Helmholtz hypothesis[21]. As summarized in 1967 by Harper[24], this fails to explain many experimental observations; for instance, that charging can occur when two pieces of the same material are rubbed against each other. While work function differences matter, alone they do not explain triboelectricity. For many decades the missing terms in triboelectricity have been unknown, despite numerous applications of the phenomenon.

We have recently[31] argued that the missing term is the flexoelectric effect, i.e., the coupling of strain gradients during asperity contacts and polarization, and subsequently extended it to include analysis of the local band bending[32]. The circumstantial evidence is strong that at a qualitative to semi-quantitative level, flexoelectricity plays a major role in triboelectricity, as our approach semi-quantitatively explains more experimental results than other approaches. To move the field forward requires rigorously testing the connection between flexoelectricity and triboelectricity, performing a quantitative experimental and theoretical analysis of the band bending during asperity contact including both the electromechanical terms and other established terms such as work function differences and semiconductor depletion regions, all without unknown parameters.

Such an analysis has many components, some of which are already partially understood. Since the early works of Terris *et al.*[33] and Jeffery *et al.*[34] it has been known force applied using a conductive atomic force microscope (CAFM) probe changes the local electronic structure. When the two materials are a metal and semiconductor, in 1953 Vick[35] pointed out the importance of Schottky barriers for triboelectricity, while in 1967 Harper[24] analyzed the triboelectric importance of the band bending of metal/semiconductor contacts. There are potentials created by the strains of a spherical indentation, which have been analyzed for piezoelectric materials[36], and more recently for flexoelectric[31,32,37,38] materials. Other aspects have been analyzed such as the variation of the effective barrier for an indentation Schottky diode as measured by Sun *et al.*[38]. A variety of works[39-42] have calculated the flexoelectric polarization or field and noted their importance in triboelectric contacts; for instance, sliding Schottky energy-harvesting devices are common[8,43-46]. Despite extensive efforts, no work has been able to develop a complete theory that quantitatively agrees with experiment for the flexoelectric case. (It should be remembered that only a small number of materials are piezoelectric, and even then in materials such as quartz, piezoelectricity is not important for triboelectricity[24].) A review of the current state of flexoelectric AFM tip pressing experiments has recently been published that covers many aspects of these experiments[47]. However, as will become apparent herein, the problem is more complex than the existing literature analysis with important terms such as the mean inner potential contributions not included in any of the prior analyses.

In this note, we develop a detailed band-bending model that quantitatively agrees with the force-dependent current-bias behavior in a CAFM Schottky diode between a $Pt_{0.8}Ir_{0.2}$ (PtIr) tip and Nb-doped strontium titanate (STO). The model follows conventional semiconductor analyses[48,49], taking into account depletion layers and image forces. A Hertzian model[50] for strains and strain gradients is used for flexoelectric and strain-dependent terms that consider the shift of the mean inner potential. Simple but appropriate models for transport are used to connect the band-bending to the experimental data in both forward and reverse bias. The analysis contains only two adjustable parameters, namely the ideality factor of the diode and the height of the Schottky barrier formed by the PtIr and STO. These are established parameters, which we fit from the experimental data. All other relevant parameters such as the elastic constants, flexoelectric coefficients, and the strain derivative of the mean inner potential are either well known or are calculated ab-initio. Based upon the analysis we can also explain how the trend of current transfer in sliding Schottky generator experiments[43,51] is surprisingly weakly dependent upon exact details of flexoelectric coefficient signs as well as shear versus tensile contributions during contacts. We argue that the experimental verification of the theoretical analysis very strongly indicates a significant if not dominant role for flexoelectricity in triboelectricity.

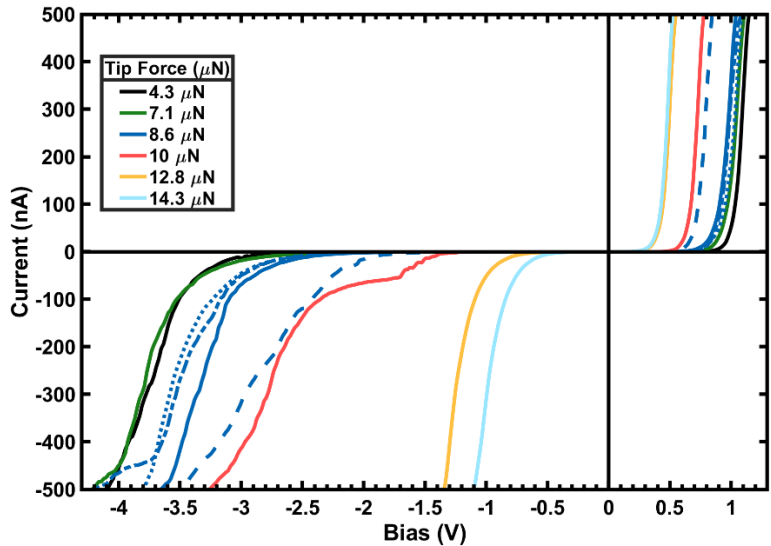

**Fig. 1.** Experiemental $I-V$ data for a $Pt_{0.8}Ir_{0.2}$ AFM tip contacting an 0.7% Nb-doped $SrTiO_3$ sample. Colors indicate different forces, with the current generally increasing towards no bias. Four repeated, but non-consecutive, measurements we collected at each tip force; the line styles

for the 8.6 µN data represent a typical set. For clarity, the remaining repetitions and some tip forces are omitted (see the Supporting Information for complete data).

Figure 1 shows how the $I - V$ characteristics of the PtIr-STO system varied with applied force $F$ for both forward and reverse bias. Increasing force shifted the bias at which a significant current was observed in all cases.

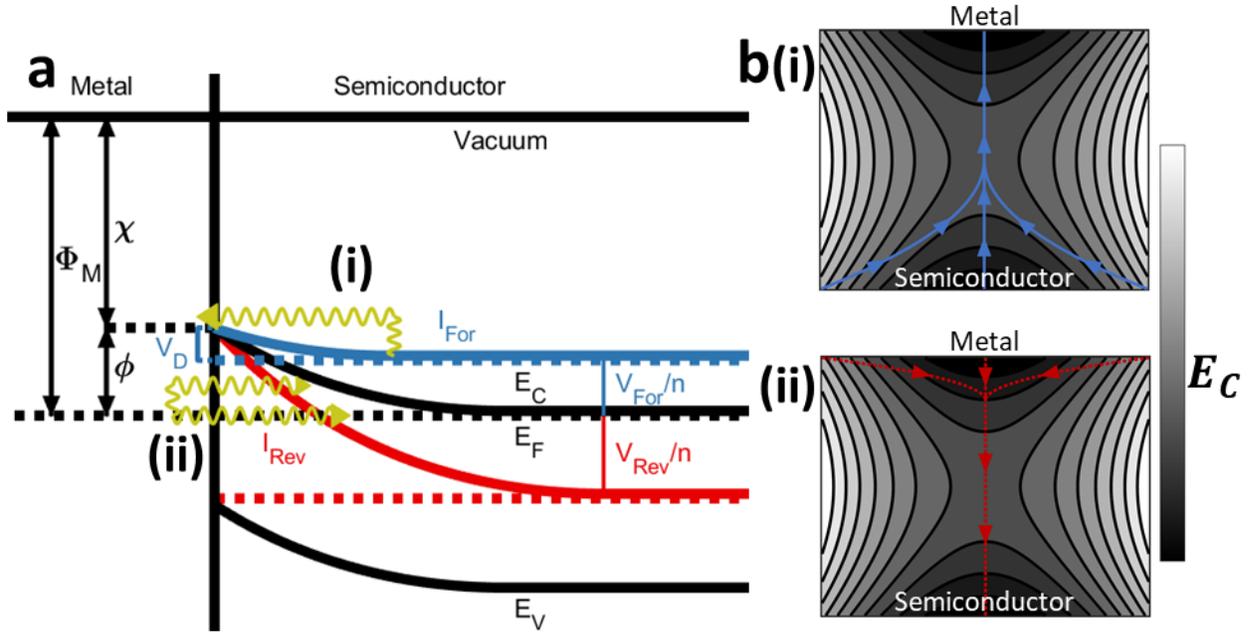

**Fig. 2.** Band-bending diagram and electron path schematics. **a**, A band diagram corresponding to the experimental setup under no strain. The metal tip is on the left and the semiconductor is on the right. The barrier height $\phi$ is the difference between the metal work function $\Phi_m$ and the semiconductor electron affinity $\chi$. $E_c$, $E_F$, and $E_V$ are the conduction band minimum, Fermi level, and valence band maximum, respectively. Black lines show the system under no bias, and the blue and red lines show the system under a forward bias $V_{For}$ and a reverse bias $V_{Rev}$, respectively. $V_D$ is the depletion potential. The valence bands are omitted because we are working with a n-type semiconductor. (i) When a forward bias is applied, a current $I_{For}$ is produced by thermally excited electrons crossing into the metal over the barrier. (ii) When a reverse bias is applied, a current $I_{Rev}$ is produced by thermally-assisted tunneling of electrons from the metal. **b**, The key point in conduction occurs at a saddle point in the conduction band minimum $E_c(F, r)$. As shown schematically, electrons may take different paths through the space-varying $E_c$, starting at different points and (i) crossing over or (ii) tunneling through the same barrier – a variational calculus problem. In both cases, the current follows paths that (i) cross over or (ii) tunnel through the inflection point.

We will first analyze the forward bias. Without any force, the system behaves as a Schottky diode with the Fermi level of the semiconductor pinned by defect states, where the pinning is with respect to the metal. (For completeness, an analysis with the pinning a fixed energy difference with respect to the semiconductor states can be found in the Supporting Information, and the results are not close to fitting the experiment.) The band structure of the zero-force case is illustrated in Fig. 2. Under forward bias, electrons in the semiconductor enter the metal by thermionic emission over a barrier of height $\phi$. The Schottky-Mott barrier height, unmodified by strain, defects, interfacial charge redistribution, or external potentials, is $\phi_0 = \Phi_m - \chi$, where $\Phi_m$ is the work function of the metal and $\chi$ is the electron affinity of the semiconductor[52,53]. When force is applied to the tip, there is electromechanical band bending which shifts the valence/conduction bands.

For forward bias using standard thermionic emission theory[49] and electromechanical band-bending, with the assumption that $V \gg k_B T/q$, the current as a function of both bias and applied force F is given by

$$I_F(V, F) = A(F) B^* T^2 \exp\left(-\frac{q\phi}{k_B T}\right) \exp\left(-\frac{q\left(V/n + V^{EM}(F)\right)}{k_B T}\right) \quad (1)$$

where $A(F)$ is the contact area, $B^*$ the Richardson constant, $T$ the temperature, $q$ the electron charge, $k_B$ the Boltzmann constant, $V^{EM}(F)$ the potential due to electromechanical terms, and $n$ the ideality factor. To interpret the experimental results, it is convenient to substitute $\phi^{eff}(F) = \phi + V^{EM}(F)$.

The area $A(F)$ was modeled by a sphere-half-space Hertzian contact[50] as

$$A(F) = \pi \left(\frac{3RF}{2E^*}\right)^{2/3} \quad (2)$$

where $R$ is the radius of the sphere (AFM probe tip) and $\frac{1}{E^*} = \frac{1}{2}\left(\frac{1-v_{STO}^2}{E_{STO}} + \frac{1-v_{PtIr}^2}{E_{PtIr}}\right)$. We use for the Young's moduli and Poisson's ratios[54,55]: $E_{STO} = 270$ GPa, $v_{STO} = 0.24$, $E_{PtIr} = 230$ GPa, and $\nu_{\text{PtIr}} = 0.37$. Due to the range of applied forces $F$, typical values of $A(F)$ range from $300 - 900$ nm$^2$ for the AFM experiments described herein.

With $A(F)$ determined, $\phi^{\text{eff}}(F)$ and the ideality factor $n$ were directly fit to each forward-bias $I - V$ curve using equation (1). We assumed $B^* = 156$ A cm$^{-2}$ K$^{-2}$ for Nb:STO samples[56], and the experiments were performed at room temperature, $T = 295$ K. Details of the fitting process appear in the Supporting Information. By plotting $\phi^{\text{eff}}$ as a function of the force, Figure 3a condensates the forward bias data presented in Figure 1 into a manageable form. Figure 3b does the same for the reverse bias data and is discussed later. Fitting $\phi^{\text{eff}}$ gave $\phi^{\text{eff}}(F) = \phi_0 + bF^{1/3}$ with $\phi_0 = 1.31 \pm 0.10$ eV and $b = -0.38 \pm 0.05$ eV/μN$^{1/3}$ (95% confidence intervals), plotted in Fig. 3a. This Hertzian-type $F^{1/3}$ scaling matches previous calculations[31] and experimental observations of tribocurrents[57]. The average value of the ideality factor $n$ was $2.04 \pm 0.09$ (95% confidence interval). Ideality factors of 1-2 are known to be appropriate depending upon carrier details[49], values of $n = 1.2 - 1.8$ have been observed[58] for progressively decreasing interface quality, and an AFM tip contact is expected to be a low quality interface. The intercept, $\phi_0$, is the unmodified barrier height and, in the Schottky-Mott limit, is the difference between the PtIr work function $\Phi_m$ and the unmodified STO electron affinity $\chi_0$. For STO, the electron affinity is approximately[59] $\chi_0 = 4$ eV. Though we find no reports of $\Phi_m$, it is bounded by[60-62] $\Phi_{\text{Pt}} = 5.6$ eV and $\Phi_{\text{Ir}} = 4.65$ eV. Using a simple linear interpolation, $\Phi_m = 0.8\Phi_{\text{Pt}} + 0.2\Phi_{\text{Ir}} = 5.4$ eV. From this, we should expect the intercept of the fit curve to be 1.4 eV, close to the experimental result.

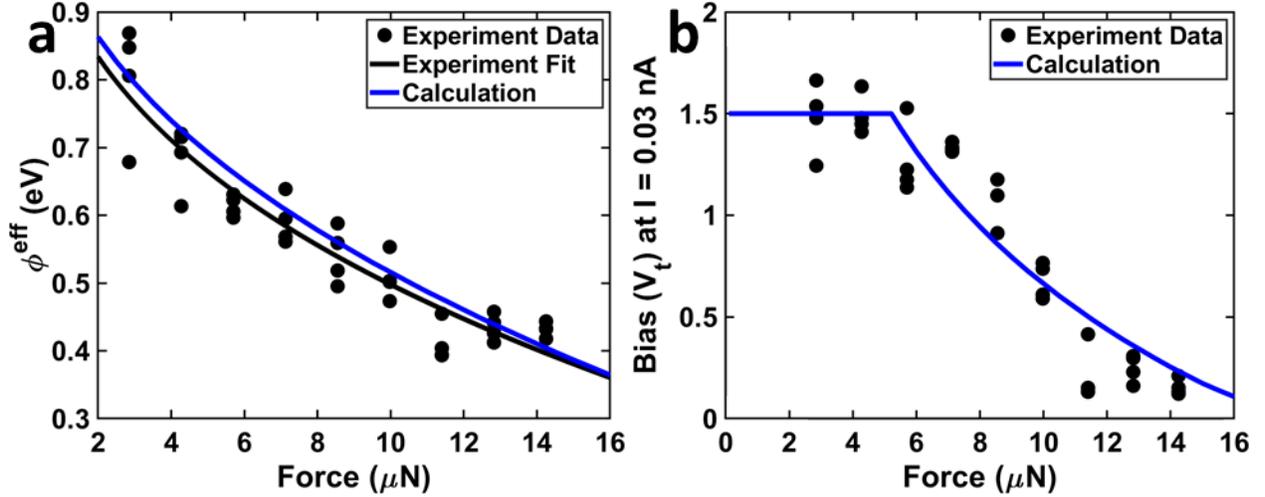

**Fig. 3.** Comparison of the experimental and calculated effective barriers (forward) and threshold biases (reverse). **a,** Experimental effective barrier heights $\phi^{\text{eff}}$ plotted against the tip force $F$ (dots), with a curve fit (black line) of the form $\phi_0 - bF^{1/3}$, with $\phi_0 = 1.31 \pm 0.10$ eV and $b = -0.38 \pm 0.05$ eV/μN$^{1/3}$ and the numerically calculated barrier (blue line). **b,** Experimental (black dots) and calculated (solid blue line) threshold biases $V_{\text{Rev,t}}$ at a threshold current of $I = 0.03$ nA in reverse bias. The experimental data suggests a breakdown bias of 1.5 V (dotted blue line).

We now turn to modelling the effective barrier $\phi^{\text{eff}}(F)$ with force $F$. Assuming cylindrical symmetry, we consider the conduction band energy $E_c(F, r)$ as a function of applied force and position $\mathrm{r} = (\rho, z)$. Assuming that any metal work function shifts due to strain are small relative to the other effects[63], then

$$E_c(F, \mathbf{r}) - E_F = V^{\text{EM}}(F, \mathbf{r}) + q\Phi_{\text{DEP}}(\mathbf{r}) + q\Phi_{\text{IMG}}(F, \mathbf{r}) \quad (3a)$$

$$V^{\text{EM}}(F, r) = \frac{dE_c}{d\varepsilon}\varepsilon(F, r) + q\Phi_{\text{FXE}}(F, r) \quad (3b)$$

Equation (3b) has two electromechanical terms on the right: the first is the deformation potential and average Coulomb potential shift due to strain (which includes "surface" effects as described later[64]) and the second is the flexoelectric potential that develops due to the strain gradient. The other terms on the right in equation (3a) are conventional diode contributions: the second term is the depletion potential that develops in the doped STO and the third the image potential on an electron crossing the semiconductor-metal interface. Since STO is centrosymmetric, we do not include any piezoelectric terms beyond the surface contributions that are already included.

We use analytical solutions of Hertzian contact theory for a PtIr sphere with radius $R$ indenting an STO half-space to calculate the strains and strain gradients necessary to quantitatively analyze equation (3). First, we consider the effects of strain without any strain gradients, which can be split into two terms, a volumetric strain and the remaining deviatoric shear strain. While volumetric strains modify the conduction band level uniformly, the shear strains cause band splitting[65], lifting degeneracies. We will consider only the volumetric strains $\varepsilon_{vol}$ and assume that the band splitting cancels on average in the experiments. The quantity of interest is then

$$\frac{dE_c}{d\varepsilon} \approx \frac{d}{d\varepsilon_{vol}}(E_c - \overline{V}) + \frac{d}{d\varepsilon_{vol}}(\overline{V} - E_{vac}) = D_{BS}^C + \varphi \tag{4}$$

where $\overline{V}$ is the average Coulomb potential in the crystal and we split $\frac{dE_c}{d\varepsilon}$ into two known derivatives. The first, the strain-induced change in the conduction band level, is the conduction band-specific deformation potential $D_{BS}^C$, as described and calculated by Stengel[66]. This term describes the shift of the conduction band edge with respect to $\overline{V}$, so we must also consider how the mean inner potential, $\overline{V} - E_{vac}$ changes with strain. The important mean inner potential term has often been called a "surface flexoelectric" contribution[64,66], but we prefer an interpretation consistent with electron diffraction as discussed in detail recently[64]. This is the second term in equation (4), $\varphi$, and is determined analytically using the Ibers approximation[64,67-69], which is accurate to ~10%. Thus, we have

$$\frac{dE_c}{d\varepsilon} \approx D_{BS}^C + \varphi = -17.2 \text{ eV} + 22.2 \text{ eV} = 5 \text{ eV} \tag{5}$$

Returning to equation (3), we next consider the bulk flexoelectric response. The polarization $P$ due to the strain gradient is given by

$$P_i(F,r) = \mu_{ijkl}\frac{d\varepsilon_{kl}(F,r)}{dx_j} \quad (6)$$

where $\mu_{ijkl}$ is the flexoelectric tensor. We assume cubic symmetry (and switch to Voigt notation), with the values[70] for STO: $\mu_{11} = -36.9$ nC/m, $\mu_{12} = -40.2$ nC/m, and $\mu_{44} = -1.4$ nC/m. The potential is therefore

$$\Phi_{\text{FXE}}(F,r) = \frac{1}{4\pi\epsilon_S}\int_\Omega \frac{P(F,r')\cdot(r-r')}{|r-r'|^3}dr' \quad (7)$$

with $\epsilon_S = 365\epsilon_0$ the STO dielectric constant[71].

Beyond electromechanical effects, the applied bias and the formation of a depletion region near the surface of doped STO must both be accounted for in $\Phi_{\text{DEP}}$. We assume $\Phi_{\text{DEP}}$ follows the form given for Nb-doped STO by Yamamoto[71], depending only upon the distance $z$ from the surface:

$$\Phi_{\text{DEP}}(r) = \frac{\sqrt{ab\epsilon_0}}{qN}\left\{\cosh\left[\cosh^{-1}\left(1 + \frac{qN}{\sqrt{a}\,b\epsilon_0}V_d\right) - \frac{qN}{b\epsilon_0}z\right] - 1\right\} + (V/n - (E_F - E_c)) \quad (8)$$

where $N$ is the dopant concentration, $a$ and $b$ parameterize the field ($\mathcal{E}$) dependence of the permittivity of STO as $\epsilon_{\text{STO}} = b\epsilon_0/\sqrt{a+\mathcal{E}^2}$ and have values $1.64\cdot 10^{15}$ V$^2$/m$^2$ and $1.48\cdot 10^{10}$ V/m respectively at room temperature[71], $V_d = \phi_0 + E_F - E_c - V/n$ is the diffusion potential, and $V$ is the applied potential. We included the image charge potential, $\Phi_{\text{IMG}}$, generated as the electrons approach the surface of the semiconductor; however, it is very small relative to the other terms (details appear in the Supporting Information).

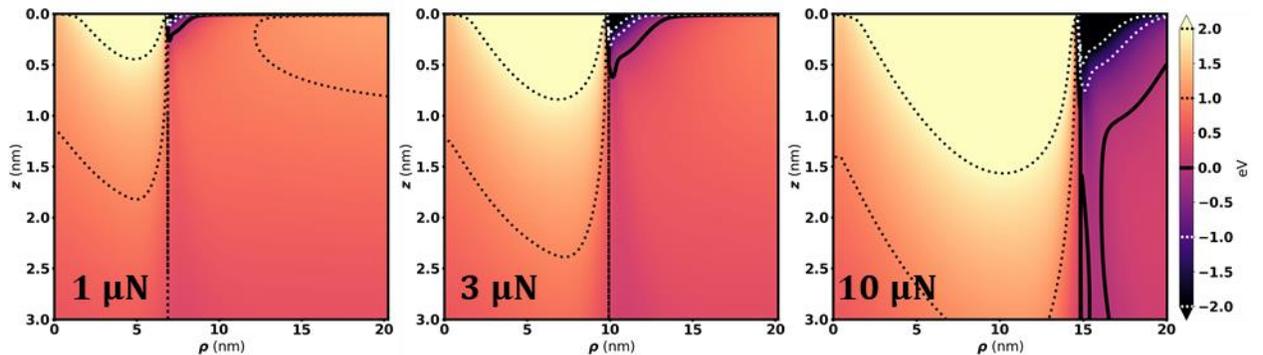

**Fig. 4.** Calculated conduction band minima for different forces. Calculated $E_c(F, r) - E_F$ for tip radius $R = 60$ nm and three different forces using cylindrical polars with $z$ into the STO and $\rho$ radial. 0 eV marks the unstrained bulk conduction band minimum. Contour lines are marked on the color bar.

Therefore, we have for $E_c$:

$$E_c(F, \mathbf{r}) - E_F = \left(D_{BS}^C + \varphi\right) + \frac{1}{4\pi\epsilon_S} \int_\Omega \frac{P(F,\mathbf{r}')\cdot(\mathbf{r}-\mathbf{r}')}{|\mathbf{r}-\mathbf{r}'|^3} \, d\mathbf{r}' + \Phi_{\text{DEP}}(\mathbf{r}) \tag{9}$$

The total band bending (radially symmetric) for the PtIr-STO system is shown in Fig. 4 for three representative forces. To compare with the experimental results, $\phi^{\text{eff}}$ was then calculated numerically by considering the paths an electron may take through the STO to the interface – a variational calculus problem. For a given path, the barrier is the maximum value of $E_c$ along that path. Then, there is a non-uniform barrier, a function of $\rho$, which is the minimum of the barriers of the possible paths to that point at the interface. Finally, a single-valued $\phi^{\text{eff}}$ was calculated from this non-uniform barrier. These calculations indicate that the current flows through a thin ring just inside the edge of contact. Further details of the calculation appear in the Supporting Information. The results for the effective barrier are shown in Fig. 3a and are in excellent agreement with the experimental data.

Turning next to the reverse bias case, we considered a reverse threshold bias $V_{\text{Rev},t}$ at which the current reaches some threshold magnitude, $|I(V_t)| = I_t$ (see Fig. 3b). The value $I_t = 0.03$ nA was just large enough to avoid instrumental noise; as seen in the Supporting Information, this localizes the current to a small region and reduces the range of different tunneling barrier heights that contribute significantly. We calculate the current by integrating for fixed radial values across the thermally-assisted tunneling barrier defined by the maximum of the STO conduction band and the metal Fermi level. This neglects the charge-transfer that can occur with electrons moving from the metal into parts of the STO very close to the tip. The current density is[72]

$$J_{\text{Tun}}(V,F,\rho) = \frac{B^*T}{k_B} \int_0^{\phi_{\max}^{\text{eff}}} P_{\text{Tun}}(E_m,V,F,\rho) \ln\left[\frac{1+\exp(-(E_m+\xi_F)/k_BT)}{1+\exp(-(E_m+\xi_F+qV)/k_BT)}\right] dE_m \quad (10)$$

where $P_{\text{Tun}}(E_m,V,F,\rho)$ is the tunneling probability for an electron with energy $E_m$ where $E_m = 0$ is the metal Fermi level. We note that this is a general form that includes thermally assisted tunneling. Using the WKB approximation[73],

$$P_{\text{Tun}}(E_m,V,F,\rho) = \exp\left[-2\int_C \sqrt{\frac{2m^*}{\hbar^2}(\phi^{\text{eff}}(V,F,\mathbf{r}) - E_m)}\, d\mathbf{r}\right] \quad (11)$$

where $m^*$ is the STO conduction band effective mass and $C$ is a path from the interface at radius $\rho$ to a point in the STO far away (where $E_c$ is constant). Of the possible paths, that with the highest tunneling probability determines $P_{\text{Tun}}$ – another variational calculus problem.

The experimentally observed current at the threshold applied bias $V_{\text{Rev,t}}$ and tip force $F$ is given by equation 12.

$$I(V_{\text{Rev,t}},F) = I_t = \int_0^{\rho_{\max}} J_{\text{Tun}}(V_{\text{Rev,t}}/n, F, \rho) 2\pi\rho\, d\rho \quad (12)$$

By restricting the analysis to low currents, the thermally-assisted tunneling is almost exclusively through the lowest part of the barrier near the edge of the contact region, the same location of current flow calculated in the forward bias case. At higher currents, the tunneling occurs over a larger region, as detailed in the Supporting Information.

A plot of $V_{\text{Rev,t}}$ with respect to the force $F$ is shown in Fig. 3b. The calculations and the experimental data agree well except for reverse biases above 1.5 V. This can be attributed to force-independent breakdown: in addition to the system missing the guard rings of modern Schottky devices[74], the sharp radius of curvature of the interface and the relatively high doping level of the STO both suggest breakdown is possible[75].

The experimental results and modelling herein are in excellent agreement, well within the experimental limits. The analysis only contains two unknown parameters, namely the ideality of the diode in forward and reverse bias and the value for the Schottky barrier between the PtIr tip and the STO substrate. All other relevant terms in the model, such as the elastic constants, flexocoupling voltage and strain derivative of the mean inner potential are either well known or are calculated ab-initio. For any specific system, the details depend on many terms, including the flexoelectric terms, temperature, elastic constants, and differences in work function. The Supporting Information discusses the contributions of the strain gradient and polarization components, and the Supporting Information discusses the contributions of the terms in equations (4), (7), and (8).

One point of some importance in a general sense is that we obtained good agreement between experiment and calculations without having atomically ordered and clean surfaces. This is because the barriers for both forward and reverse bias are beneath the surface by $1-3$ nm (see the Supporting Information). This means that the behavior of a particular material will be somewhat consistent, which is what has been found experimentally. Of course, chemisorbed polar molecules or water will have an effect on the zero-force barrier, but they will not change that much the band-bending change due to the flexoelectric effect.

Though the calculations do not explicitly address charge transfer, they offer some insights that agree with experiments for sliding Schottky generator experiments[43,51] that observe electron transfer from a semiconductor to a metal, surprisingly independent of the sign of the flexoelectric coefficients and details of shear versus indentation or plowing. What one has is a type of ratchetting charge pump. More details are provided in the Supporting Information; we will summarize here and in Fig. 5. Near the contact there is a region where the bands bend down in the semiconductor as force is applied, these regions will pull in electrons from the body of the semiconductor far from the contact region. As the force is released there is automatically a barrier for these electrons to move back into the semiconductor, so transfer to the metal is favored. Changing the sign of the flexoelectric coefficients will change whether this downwards region is towards the outside or the inside of the circular contact region, but not its presence. Similarly, details of shear versus compression or other types of asperity contact will change the exact magnitude of the region where the band bends significantly down, but the overall details will remain the same.

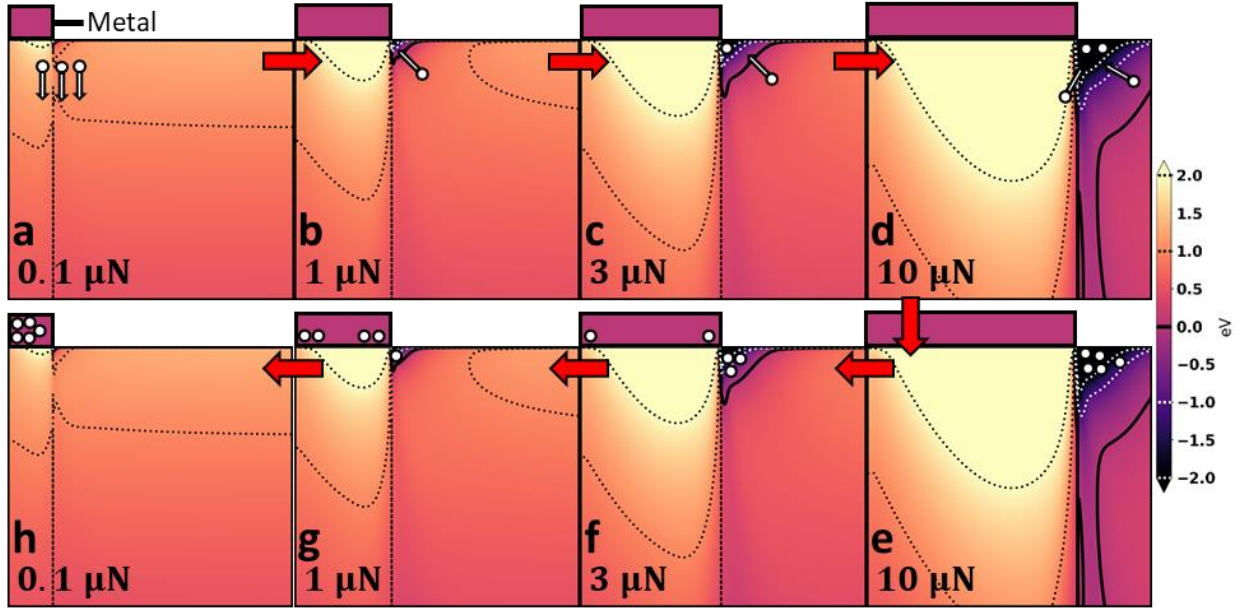

**Fig. 5.** Ratcheting mechanism for charge transfer in metal-semiconductor contacts. **a-h**, $E_c(F,r) - E_F$ in the range $0 \leq \rho \leq 20$ nm and $0 \leq z \leq 3$ nm for increasing and decreasing forces $F$, as in contact and pull-off of an asperity. **a**, For very low forces, the depletion potential is most important. **b-d**, As the force increases, electrons (white circles, schematically arrowed) move from regions with an increasing potential to regions with decreasing potential. **e-g**, As the force decreases, the number of available states in the potential well decreases, forcing some electrons into the metal rather than back across the barrier in the semiconductor, where states near the metal Fermi level $E_{F,M} = 0$ are available. **h**, After the force is completely released, electrons have transferred to the metal.

We have presented a model of electromechanical band-bending in a triboelectric metal-semiconductor system that quantitatively agrees with the force-dependent current-bias experiments performed herein and qualitatively agrees with charge transfer observed in other works. This offers insight into the details of the triboelectric effect in metal-semiconductor systems and provides strong evidence of the importance of flexoelectric band-bending to triboelectricity. Rather complex mechanics, band-bending physics and device physics combine to end up producing a relatively simple and quite general result. While details are of course very strongly dependent upon specifics of the materials, broad-brush results for charge transfer are surprisingly invariant of, for instance, the amount of shear at the contact. If this was not the case, then the simple concepts such as the 19th century triboelectric series would never have had even enough limited success such that it could be taught to generations of high-school students.

Is triboelectricity as a ratchetting charge pump generalizable from the Schottky contacts herein to other cases? Perhaps.

**Acknowledgments:** The authors acknowledge financial support from the National Science Foundation (NSF) under grant number DMR-1507101 and the U.S. Department of Energy, Office of Science, Basic Energy Sciences, under Award No. DE-FG02-01ER4594.

**Author contributions:** K. P. O. performed the experiments and analysis with advice from C. A. M. and supervision from L. D. M. All authors contributed to the writing of the Letter.

**Competing interests:** The authors declare no competing financial interest.

# Supporting Information for

## Band bending and ratcheting explain triboelectricity in a flexoelectric contact diode

K. P. Olson, C. A. Mizzi, L. D. Marks[*]

Department of Materials Science and Engineering, Northwestern University; Evanston, Illinois, United States of America.

*Corresponding author. Email: l-marks@northwestern.edu

**Supplementary Notes**

**SN1: Methods**

To quantify the band bending, we used a conductive atomic force microscope (CAFM) with solid $Pt_{0.8}Ir_{0.2}$ probe tips (25PtIr300B from Rocky Mountain Nanotechnology) to collect current-bias ($I - V$) curves at various tip-sample contact forces, $F$. The samples were 0.7 wt% Nb-doped (100) STO single crystals from MTI Corporation, with dimensions approximately 10x10x0.5 mm. The bottoms of the samples were electrically connected to the conductive sample stage with silver paste. Before CAFM measurements, samples were cleaned in acetone and ethanol. CAFM measurements were performed in contact mode CAFM with a Bruker Dimension Icon AFM.

To minimize artifacts, as discussed further in the Supporting Information, voltages were limited to avoid resistive switching (Supporting Information), data with large ideality factors and other artifacts due to bad contacts were not used (Supporting Information), and care was taken to avoid artifacts from plastic deformation of the tips (Supporting Information).

The probe tip radii, $R$, were measured with a FEI Quanta 650 Environmental Scanning Electron Microscope and were typically 60-70 nm. The tip force $F = k\Delta x$ was determined from the deflection of the cantilever $\Delta x$ and the calibrated probe spring constant $k$. Because we used stiff probes ($k = 5.7 \pm 1.2$ N/m) the spring constant was determined via the Sader method[76]; more details can be found in the Supporting Information.

**SN2: Spring Constant and Tip Radius Determination**

The spring constants of the AFM probes were fit by the Sader method[76,77], as described in the main text. This method requires only the dimensions of the cantilever (measured via optical microscopy) and the thermal power spectral density, plotted with solid lines in Fig. S1a.

The power spectral density is fitted with

$$A = A_{\text{white}} + \frac{A_0 f_0^4}{(f^2 - f_0^2)^2 + \left(\frac{f f_0}{Q}\right)^2} \tag{S1}$$

where $A$ is the amplitude (power) and $f$ is the frequency. The fit parameters are $f_0$, the resonant frequency, $Q$, the quality factor, $A_0$, the zero-frequency amplitude, and $A_{\text{white}}$, the white noise baseline. These fits are shown with dotted lines in Fig.S1a.

Ideally, only a single peak and its modes should exist. Mechanical or electrical noise from the AFM system can induce extra peaks, such as those near 16 and 32 kHz (see Fig. S1a inset). Note that the shapes and approximate location of these peaks do not change between probes, while the peaks of interest, with resonant frequencies between 20 and 25 kHz, do differ for different probes. The manufacturer's specification for the resonant frequency was 21 kHz ± 30%.

Fitting these peaks, the spring constant for each probe can be determined with Sader's result

$$k = 7.5246 \rho_f w^2 L Q f_0^2 \Gamma_i(\text{RE}) \tag{S2}$$

where $\rho_f$ is the density of the fluid (air), $w$ and $L$ are the width and length of the cantilever, and $\Gamma_i$ is the imaginary component of the hydrodynamic function of RE, the Reynold's number.

Using the online calculator provided by Sader [http://www.ampc.ms.unimelb.edu.au/afm/calibration.html] gave spring constants of about $6 - 7$ N/m, depending on the probe.

The probe tip radius was determined by inspection of SEM images. Fig. S1b shows one image with an overlayed circle that shows the tip radius was approximately $R = 60$ nm; This was confirmed with other images.

### SN3: Full I-V Data

The full set of $I - V$ data referenced by Fig. 1 is plotted in Fig. S2.

### SN4: Thermionic Emission Fitting and Measurement Details

Standard thermionic emission theory gives a current[48]

$$I(V) = AB^* T^2 \exp\left(-\frac{q\phi^{\text{eff}}}{k_B T}\right) \left[\exp\left(\frac{qV}{nk_B T}\right) - 1\right] \tag{S3}$$

where $V$ is the applied bias, $A$ the contact area, $B^*$ the Richardson constant, $T$ the temperature, $q$ the electron charge, $k_B$ the Boltzmann constant, $\phi^{\text{eff}}$ the effective barrier, and $n$ the ideality factor. For $V/n \gg k_B T/q \approx 0.025$ V at room temperature, this gives

$$I(V) = AB^* T^2 \exp\left(-\frac{q\phi^{\text{eff}}}{k_B T}\right) \exp\left(\frac{qV}{nk_B T}\right) \tag{S4}$$

Equation (S5) then provides a convenient way to determine $n$.

$$\ln I = \ln AB^* T^2 - \frac{q\phi^{\text{eff}}}{k_B T} + \frac{qV}{nk_B T} \Rightarrow n = \frac{q}{k_B T} \frac{d \ln I}{dV} \tag{S5}$$

Typical fits for $n$ and $\phi^{\text{eff}}$ are shown in Fig. S3a and S3b, respectively, where one curve at each force is shown. Here $n$ is fit with the linear least squares method to the portion of the $\ln I - V$ data with $\ln I$ [nA] $> -1$, so that the noise floor is excluded. Using the value of $n$, $\phi^{\text{eff}}$ is then fit by a non-linear least squares method. The high quality of the fits supports our assumption of thermionic emission.

Each of these measurements are averages of at least 10 $I - V$ curves measured immediately after one another. The multiple measurements at a single value of $F$ were separated by longer times or measurements at other values of $F$.

Fig. S4 shows the ideality factor fit for each curve in Fig. S2. The average value is $n = 2.04 \pm 0.09$ (95% confidence interval). There is no trend with force, so reporting and using the average $n$ is appropriate.

# SN5: Experimental Notes

## SN5.1: Contact Quality and Resistive Switching

The $I-V$ measurements are quite sensitive to the quality of contact. Fig. S2 shows some variation even at the same location with the same force $F$. Larger variations were present in measurements performed at different locations on the sample. $I-V$ behavior that strays from perfect thermionic emission is described by the ideality factor $n$, which can be related to the quality of the metal-semiconductor interface[58]. Here, we report measurements with $n$ in the range $1.5 - 3$. Data with $n > 3$ had significant deviation from thermionic emission theory and showed much more scattered, and sometimes even non-monotonic, effective barrier-force relationships. These cases can be attributed to poor probe-sample contact, significant sample impurities near the contact region, or other non-ideal conditions; these are not shown in Fig. S2.

The measurements exhibited resistive switching, a well-known phenomenon in perovskite oxides such as STO[58,78]. This effect can be attributed to the temporary filling and emptying of states above the conduction band. Filling of these states occurs during the $I-V$ curve acquisition if the applied bias values are large relative to the barrier heights; this was therefore avoided. Values reported and analyzed in the main text and other Supplementary Note sections always consider the forward-sweeping direction, which starts with a reverse bias and sweeps towards a forward bias. An example of the observed resistive switching is shown in Fig. S5.

In addition to the switching observed during bias sweeps, we observed long-term (~ 1 hr) hysteresis when particularly large reverse biases were applied. A point on the sample was held at a reverse bias $V_h$ for 1 minute. Immediately after, a series of CAFM images were obtained at 0 bias. For large $V_h$, an increased current was observed in the images. The current was significantly different from the $V_h = 0$ V control case when $V_h \geq 5$ V and the difference was especially large when $V_h \geq 8$ V, as shown in Fig. S6a. This can be explained by charge building up near the probe-sample interface under a large $V_h$, then moving into the unbiased probe as it collects the subsequent image. Fig. S6b shows how this effect and decayed over time. These findings suggest large effects were avoided by using a bias range for $I-V$ sweeps of $-3$ to $+3$ V. We note that this long-term effect is similar to that observed by Morita[79].

## SN5.2: Check for Plastic Deformation or Other Hysteretic Effects

The data presented in Fig. 3 was collected in two passes. In the first pass, three sets of $I-V$ curves were collected at the minimum force, then at progressively increasing forces. The second pass restarted the measurements at the lowest force, where one set of $I-V$ curves were collected at increasing forces. Fig. S7 clearly shows there was no trend in the results of the second pass relative to the first pass, which suggests there was no plastic deformation of the tip during the course of the measurements or sample or any other significant hysteretic effects. Similar procedures with other data sets, including sweeping from the highest to lowest force, also showed no trends that indicated any significant hysteretic effects.

# SN6: Image Potential

We consider an electron in the semiconductor moving towards the interface with the metal. As it approaches the interface, it will feel a Coulombic image potential[80] equal to

$$\Phi_{\text{IMG}}(z) = \frac{q}{16\pi\epsilon_S z} \tag{S6}$$

which causes a contribution to the band bending of

$$q\Phi_{\text{IMG}}(z) = \frac{q^2}{16\pi\epsilon_S z} \tag{S7}$$

where $\epsilon_S$ is the dielectric constant of the semiconductor. Here, with $\epsilon_S = 365\epsilon_0$ and a minimum useful value $z = 0.1$ nm, this shift is only 10 meV, which is very small compared to the shifts due to strains and strain gradients. While this effect was included in the calculations, it is ignored in the main text because it was always small.

## SN7: Numerical Calculation of the Effective Barrier

We consider an electron that starts far away from the interface and will cross the barrier. Without corrections to be discussed later, the barrier for a particular path the electron takes to the metal tip is the maximum potential along that path. Now, we assume the electron will take the path that has the smallest barrier (this approach turns out to be very close or exactly the path that follows a gradient-based approach where the electron moves based on the potential gradient at its position). Using this approach while forcing the path to end at the position ($\rho, z = 0$) (a position at the metal-semiconductor interface), we calculate the barrier $\phi^{\text{calc},0}$ as a function of $\rho$. Fig. 2b shows a schematic of the saddle-type point (Fig. S8 shows the saddle point in the calculated $E_c(F, \mathbf{r})$ at $F = 3$ μN) that paths near the contact edge pass through (resulting in the flat minimum of $\phi^{\text{calc},0}(\rho)$ in Fig. S9). Fig. S10 shows the depth, $z$, of the saddle point as a function of the force $F$. The depth of the saddle point scales almost exactly as $F^{1/3}$, as expected for Hertzian-type contacts. This suggests the depth of the barrier is primarily dependent on the electromechanical terms.

Now we discuss the necessary corrections to $\phi^{\text{calc},0}(\rho)$. First, the analysis of the experimental data assumes an effective barrier $\phi^{\text{calc},1}$ that is constant over the contact area (a circle of radius a). Keeping the currents equal for the constant barrier $\phi^{\text{calc},1}$ and the spatially varying $\phi^{\text{calc},0}(\rho)$ cases,

$$I = \pi a^2 B^* T^2 \exp\left(-\frac{q\phi^{\text{calc},1}}{k_B T}\right) \exp\left(\frac{qV}{nk_B T}\right) = \int_0^a 2\pi B^* T^2 \exp\left(-\frac{q\phi^{\text{calc},0}(\rho)}{k_B T}\right) \exp\left(\frac{qV}{nk_B T}\right) \rho \, d\rho \tag{S8}$$

Therefore, the following expression gives $\phi^{\text{calc},1}$

$$\phi^{\text{calc},1} = -\frac{k_B T}{q} \ln\left[\frac{2}{a^2} \int_0^a \exp(\phi^{\text{calc},0}(\rho)/k_B T) \rho \, d\rho\right] \tag{S9}$$

A second correction to this $\phi^{\text{calc},1}$ is necessary to account for the finite temperature. Assuming a Fermi-Dirac distribution of electron energies and a density of states $\text{DOS} = CE^{1/2}$, where $C$ is a constant such that

$$\int_0^\infty C\sqrt{E}/\left(1 + \exp\left(\frac{E}{k_B T}\right)\right) dE = 1 \tag{S10}$$

The value after this correction, $\phi^{\text{calc},2}$, is given by

$$\phi^{\text{calc},2} = -k_B T \ln\left[\int_0^\infty \exp\left(-\frac{\phi^{\text{calc},1} - E}{k_B T}\right) \frac{C\sqrt{E}}{1 + \exp\left(\frac{E}{k_B T}\right)} dE\right] \tag{S11}$$

A third and final correction accounts for the position of the Fermi level with respect to the conduction band minimum. The Fermi level of 1-in-64 (0.80 wt%) Nb-doped STO was calculated using WIEN2k[81] using a 4x4x4 supercell with a single Nb atom substituting for Ti. Technical parameters were RMTs of 2.1, 1.9, 1.88 and 1.7 for the Sr, Nb, Ti and O respectively and an RKMAX of 6.5, using the PBEsol functional[82] as well as on-site hybrid methods we have used in previous work on STO (see [83,84] and references therein). This gave a number 65 meV above the conduction band minimum. Because $\phi^{\text{calc},2}$ is calculated with respect to the conduction band minimum, the calculated effective barrier is given by

$$\phi^{\text{eff,calc}} = \phi^{\text{calc,2}} - 65 \text{ meV} \tag{S12}$$

Fig. S8 shows the results of the process described above for $F = 3$ μN. $\phi^{\text{eff,calc}}$ is the value that can be compared to the effective barrier extracted from the experimental data and is the blue line plotted in Fig. 3.

## SN8: Numerical Calculation of Reverse-Bias Tunneling Current

Fig. S11 shows the tunneling probabilities $P_{\text{Tun}}$ and the integrand $P_{\text{Tun}} \ln\left[\frac{1+\exp(-(E_m+\xi_F)/k_BT)}{1+\exp(-(E_m+\xi_F+qV)/k_BT)}\right]$ (see equation (10)) at three different biases, which give three different currents. As the bias increases, the tunneling electrons have decreasing energy. As the bias increases further, electrons begin to tunnel away from the edge of the contact region. Two effects support using the smallest current for the analysis. First, the tunneling model does not include other sources of current such as avalanche breakdown. Additionally, as the bias increases, the tunneling electrons have increased energy above the conduction band minimum. Therefore, the density of states parameterized by the effective mass used in the model will be increasingly erroneous. To minimize both sources of error, we analyze the reverse bias current thresholds at the smallest practical current.

Figs. S11d-f clearly shows that the energy bounds of the calculations, $E_m < E_{\max} = 0.3$ eV, are sufficient to capture all the significant contributions to the current.

## SN9: Pinning

Two cases are possible, in principle, for pinned states at the metal-semiconductor barrier: pinning to states that remain fixed relative to the metal Fermi level or pinning to states fixed relative to the semiconductor Fermi level. The band bending for the two cases is shown in Fig. S12a and S12b, respectively. In the former case, the energy of the bands at the interface does not change with applied bias, while in the latter, the states shift with applied bias. Calculations of $\phi^{\text{eff}}$ in forward bias and the threshold bias $V_{\text{Rev,t}}$ in reverse-bias data were performed for both cases. The key difference in the calculations is $V_d$. For the first case, $V_d = \phi_0 + E_F - E_c - V/n$ while for the second case, $V_d = \phi_0 + E_F - E_c$. The resulting predictions are shown in Fig. S12c-d and clearly show that the case of pinning to states fixed relative to the metal appropriately matches the experimental data.

## SN10: Strain Gradient and Polarization Component Analysis

Figs. S13-S14 show the strain gradient components and polarization components for $F = 3$ μN. Figs. S15-S16 show the contributions of each of these components to $\Phi_{\text{FXE}}$, i.e., the value of $\Phi_{\text{FXE}}$ if all other components are artificially set to 0. These show that both polarization components and at least the five strain gradient components $\varepsilon_{\text{rrr}}$, $\varepsilon_{\text{zzr}}$, $\varepsilon_{\text{rrz}}$, $\varepsilon_{\theta\theta z}$, and $\varepsilon_{\text{zzz}}$, contribute significantly.

## SN11: Charge Transfer and Flexoelectric Signs

### SN11.1: Impact of the Sign of the Flexoelectric Coefficient

Fig. S17 shows a tableau of results for the flexoelectric potential $\Phi_{\text{FXE}}$ for all the permutations of flipping the signs of the flexoelectric coefficients $\mu_{11}$, $\mu_{12}$, and $\mu_{44}$. Figs. S17a and S17h exhibit the expected behavior – when all three coefficients' signs are flipped, the sign of $\Phi_{\text{FXE}}$ is flipped. Because $|\mu_{44}| \ll \mu_{11}, \mu_{12}$, the impact of flipping the sign of $\mu_{44}$ is diminished. When the signs of $\mu_{11}$ and $\mu_{12}$ are different, the potential map varies in space in a complex manner. Most importantly, however, we note that regardless of the signs of any flexoelectric coefficient, there is always a portion of the potential that is negative and in contact with the metal-semiconductor interface. Therefore, the fact that charge transfer occurs for a wide variety of materials with different (and unknown) flexoelectric coefficients[43,50] is explained by our model.

### SN11.2: Charge Transfer Analysis

The analysis and calculations ignore charge transfer, but some qualitative insight into charge transfer can be found from the calculations of the conduction band minimum $E_c$. At a metal-semiconductor interface, there are many in-gap states. The effect of the calculated band-bending on electrons in these states qualitatively explains the direction of charge transfer, and why the direction is independent of the signs of the flexoelectric coefficients.

Fig. 5 of the main text shows a schematic of the charge transfer process with flexoelectric coefficients $\mu_{11} = -36.9$ nC/m, $\mu_{12} = -40.2$ nC/m, and $\mu_{44} = -1.4$ nC/m, and Fig. S18 shows the same schematic with opposite coefficients. As the force increases during asperity contact, electrons fill states where the band bends down. Then, as the force decreases during pull-off, these states increase in energy above the metal Fermi level $E_{F,M}$ and electrons consequently transfer to the metal.

Fig. S17 shows the flexoelectric potential $\Phi_{FXE}$ for various signs of the flexoelectric coefficients. In all cases, regardless of the flexoelectric coefficients, the band bends down in parts of the semiconductor near the interface and facilitate the charge transfer described in Fig. 5 of the main text. That charge transfer occurs regardless of the flexoelectric sign (see Fig. S18) agrees with the results of sliding Schottky generator experiments[43,50], which observe current flowing from the metal for a variety of substrate materials which have different and often unknown flexoelectric properties.

A quantitative description of the charge transfer is outside the scope of our current analysis and is left to future work.

### SN12: Relative Mangitudes of the Potential Contributions

We analyze the relative importance of the contributions to the total effective barrier (see equation (3)) by splitting the calculated potential into its components, shown in Fig. S19. These calculations make clear that the primary source of the shifts is the flexoelectric potential, which is roughly an order of magnitude more impactful than the effects from strain or the depletion potential. Thus, the force dependence of the $I - V$ characteristics is explained by the barrier shift caused primarily by the strain gradient. The strain gradient increases with force as $F^{1/3}$ in Hertzian contact, so the dominance of this term explains the $\phi_0 - bF^{1/3}$ fit in Fig. 3a. Additionally, strain gradients increase with decreasing length scales. That is, this effect is most relevant at the nanoscale, where large strain gradients of $\sim 10^7 - 10^8$ m$^{-1}$ can develop from sources such as AFM probe contact, as analyzed in this work, or epitaxial strains in a thin film[85].

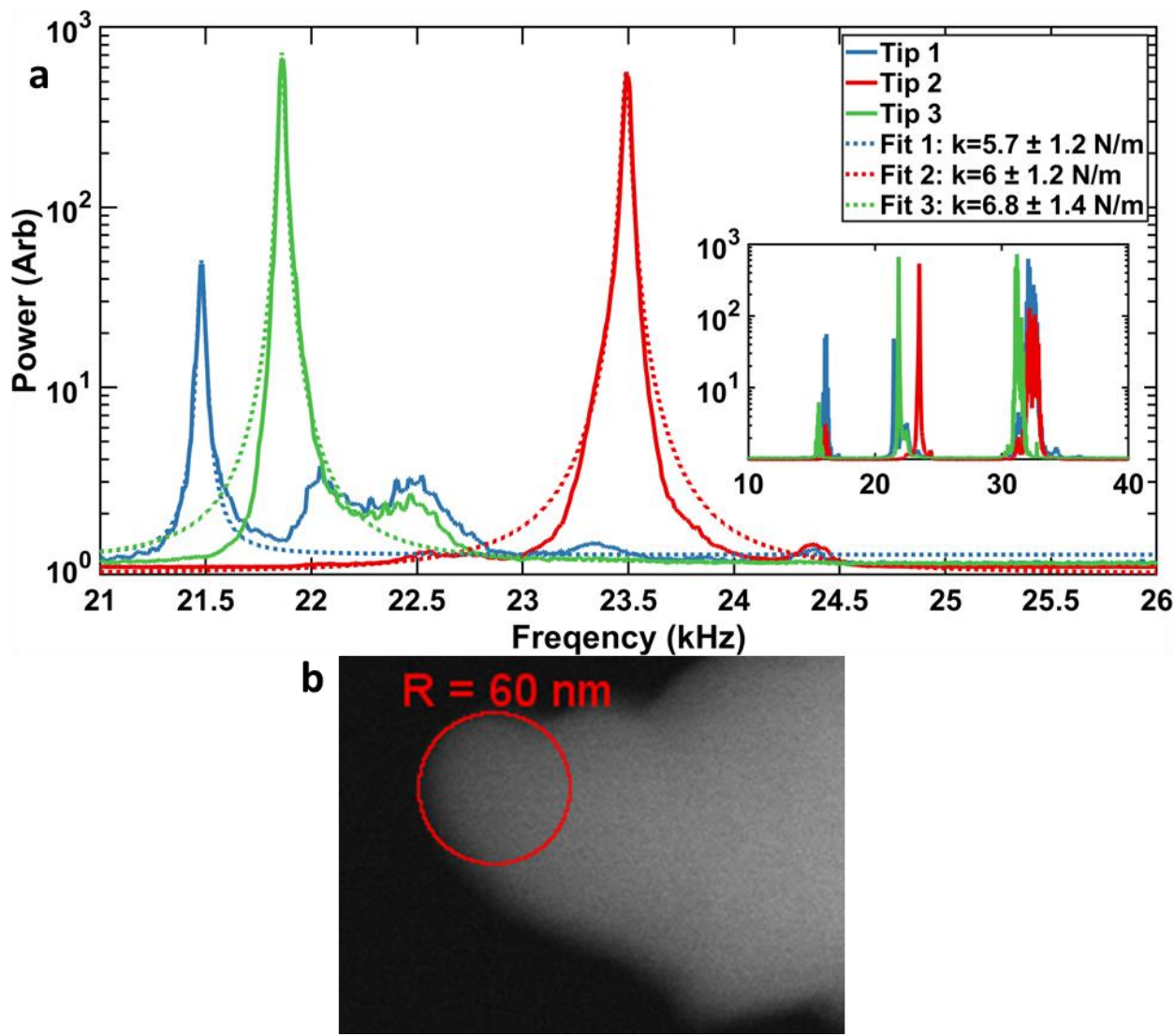

**Fig. S1.**

**a,** Power spectral density plots for three AFM probes showing the spectrum (solid) and fits (dotted) used to calculate the probe spring constants, the values of which are given in the legend. (inset) The full spectrum shows the first and third peaks, at approximately 16 and 32 kHz, which are not representative of the probes. **b**, SEM image of an AFM probe tip after use. The overlaid circle shows the tip radius is 60 nm.

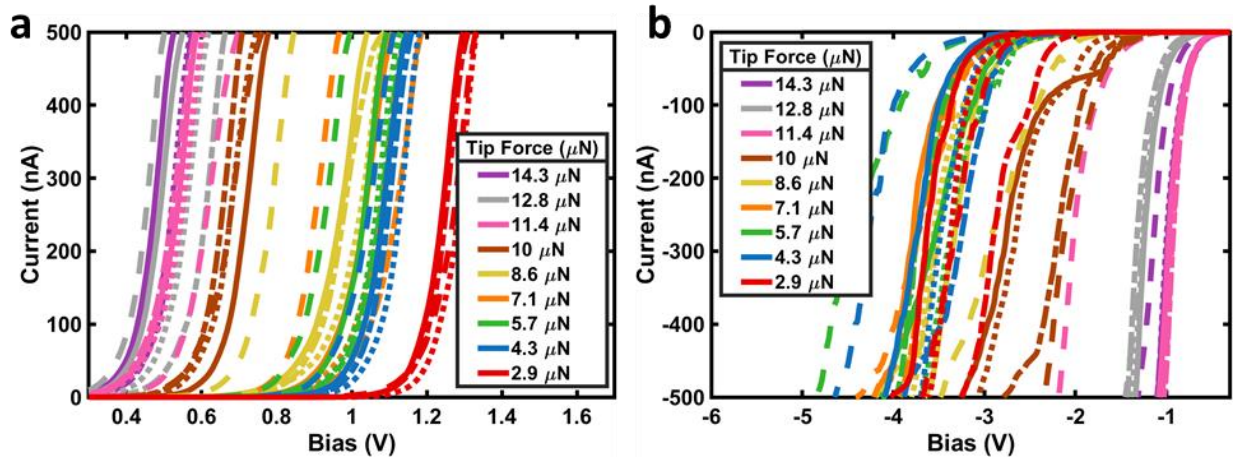

**Fig. S2.**

**a,b**, The complete set of data corresponding to that presented in Fig. 1 in reduced form; Forward (**a**) and reverse (**b**) bias $I-V$ data for a $Pt_{0.8}Ir_{0.2}$ AFM tip contacting an 0.7% Nb-doped $SrTiO_3$ sample. Colors indicate different forces, generally increasing to the left. Line styles represent repeated, but non-consecutive, measurements with the same probe and sample and at the same location on the sample.

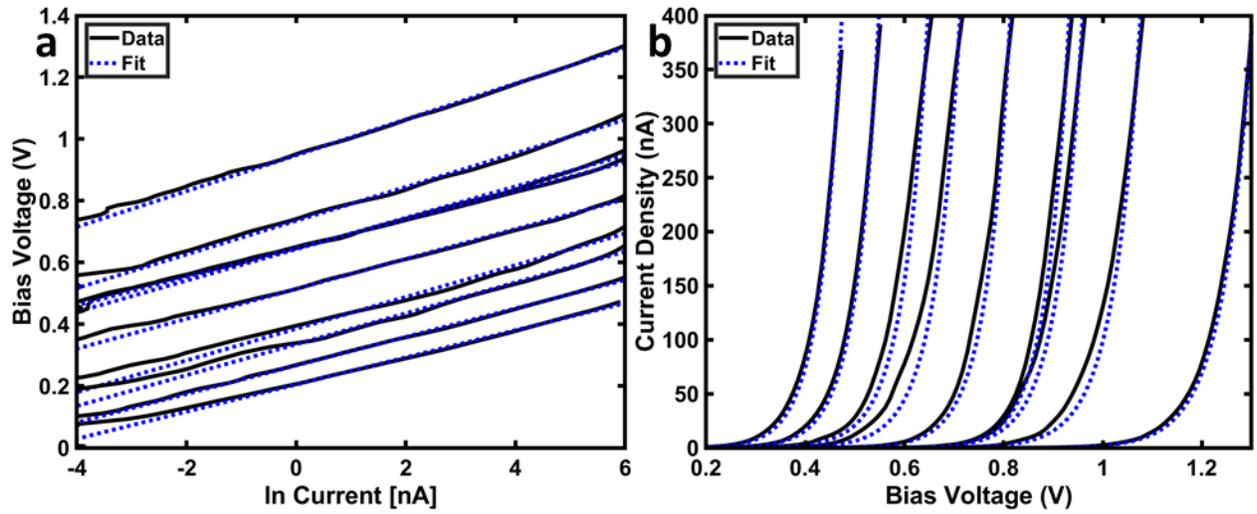

**Fig. S3.**

**a,b**, Data (solid black) and fits (dotted blue) used to determine $n$ (**a**) and $\phi^{\text{eff}}$ (**b**). Each fit shown here corresponds to a curve in Fig. 1a and a single data point on the plots in Fig. 3.

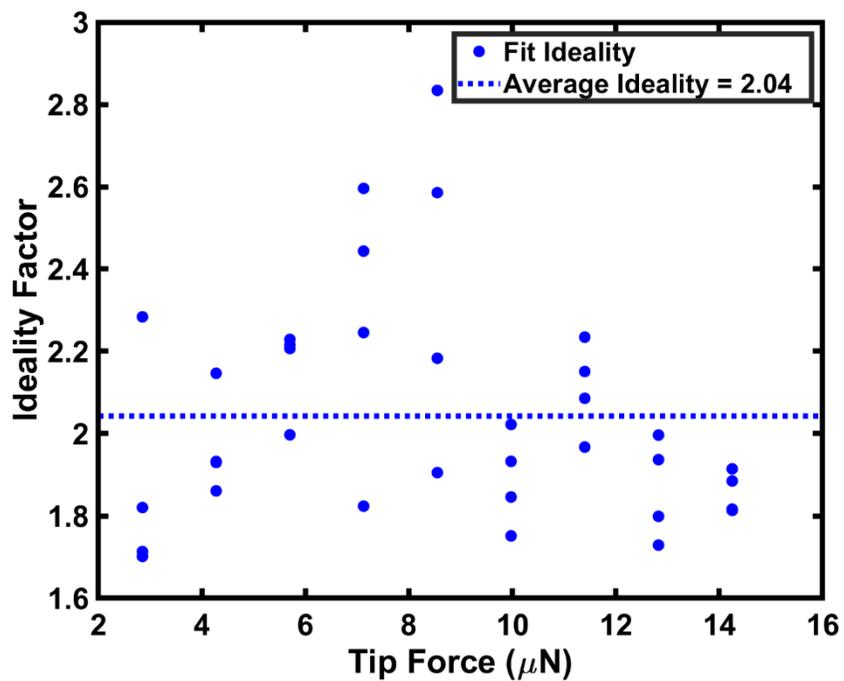

**Fig. S4.**

Ideality factor for each curve in Fig. S2. The average value is $n = 2.04 \pm 0.09$ (95% confidence interval).

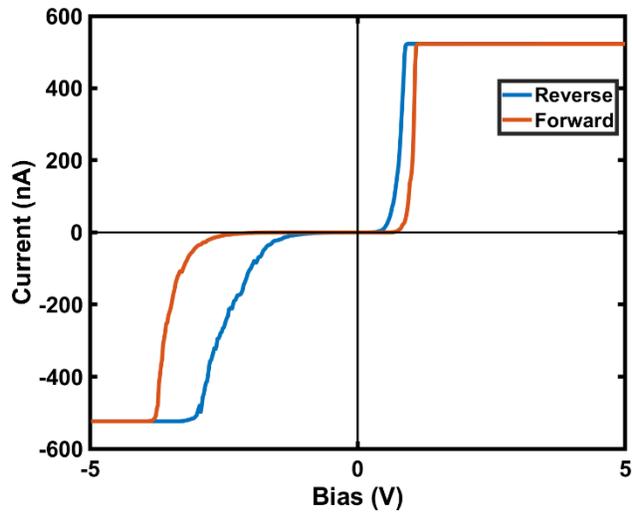

**Fig. S5.**

An example of resistive switching in $I-V$ measurements. The red curve is the initial forward-sweeping bias ($-V$ to $+V$) and the blue curve is the reverse-sweeping ($+V$ to $-V$) collect immediately after. Resistive switching for STO is well documented; we avoided all data where this occurred.

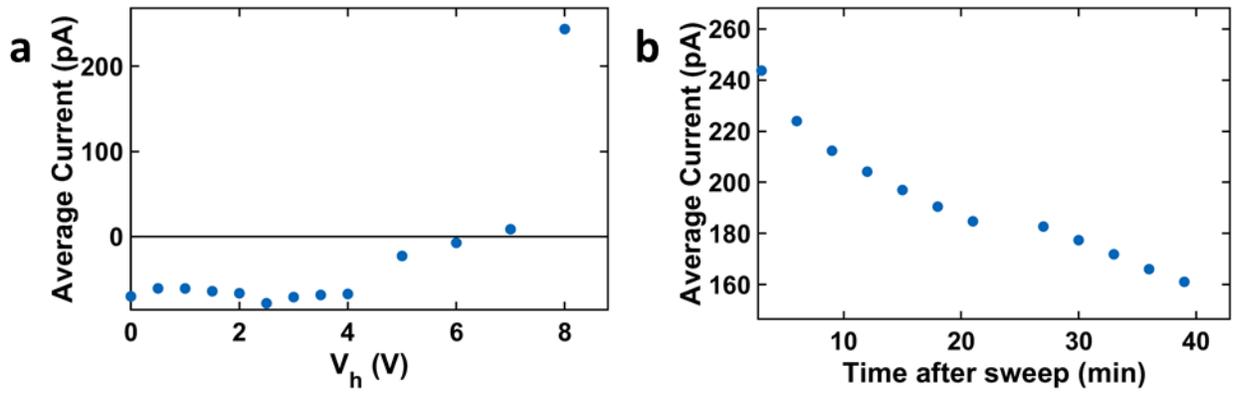

**Fig. S6.**

**a**, Average current in a 500 nm x 500 nm CAFM image collected at 0 bias, 3 minutes after being held at a reverse bias $V_h$. **b**, The same in images collected at different times after being held at a reverse bias $V_h = 8$ V. Anomalies due to charge buildup were avoided.

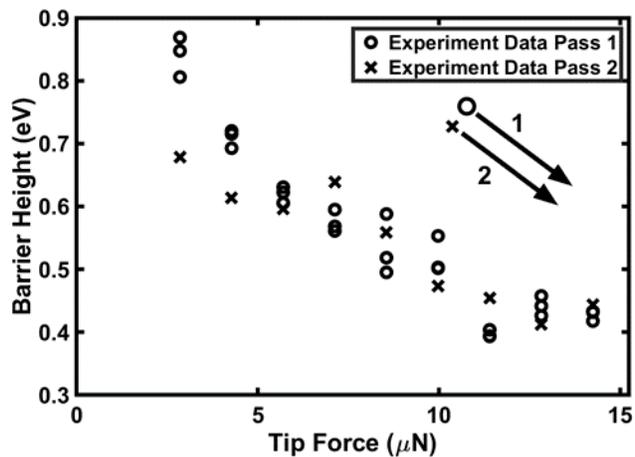

**Fig. S7.**

Barrier heights $\phi^{\text{eff}}$ from forward bias data plotted against the tip force $F$. The first pass collected three data points at each force, starting at the lowest force and increasing to the maximum force. The fourth data point was collected in a second pass immediately after, again starting at the lowest force and increasing to the maximum force. There is no consistent trend in the fourth point relative to the others, suggesting a lack of hysteretic effects.

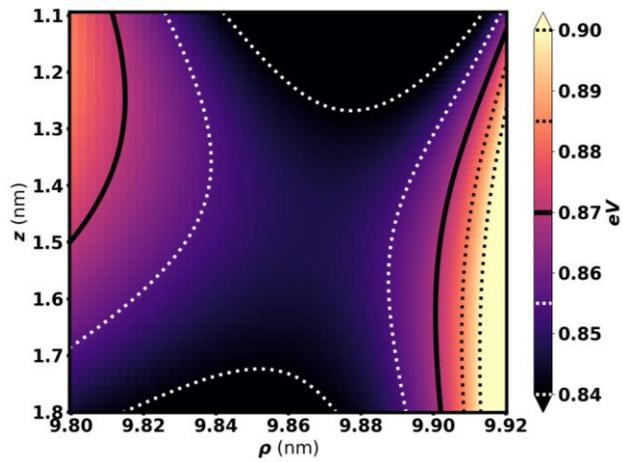

**Fig. S8.**

A saddle point in $E_c(F, \mathbf{r}) - E_F$ for $F = 3$ μN and $R = 60$ nm calculated from the model described in the text. Fig. 2b explains the importance of the saddle point and shows a schematic representation. Contour lines are marked on the color bar.

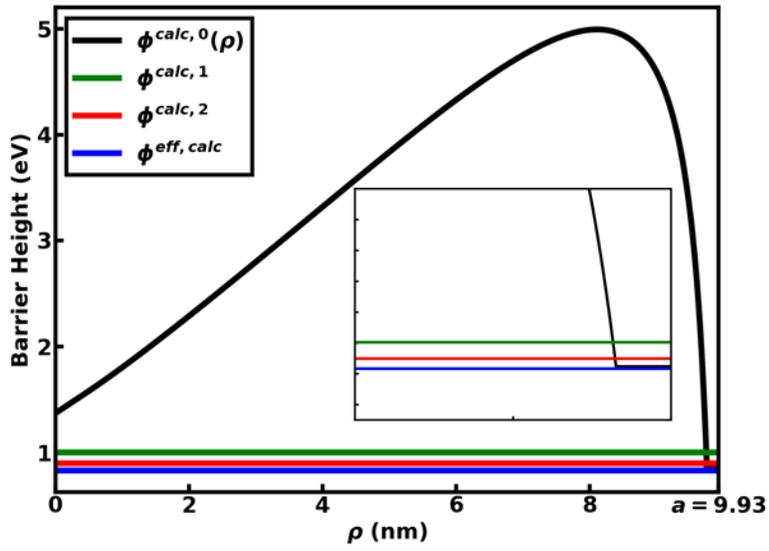

**Fig. S9.**

Plot of the non-uniform barrier height $\phi^{\text{calc}}(\rho)$ and the barrier height at other steps of the calculation. (inset) Plot zoomed on the region $0.9a \leq \rho \leq a$ and $0.5 \sim 2$ eV, where $a$ is the contact radius.

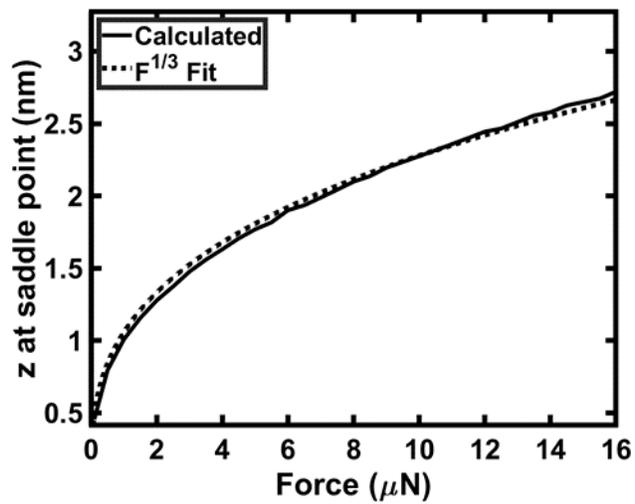

**Fig. S10.**

The depth $z$ of the saddle point in $E_c(F, \mathbf{r}) - E_F$ as a function of the force $F$. A fit of the form $z = 1.079 \text{ nm}/\mu\text{N}^{1/3} \cdot F^{1/3}$ is plotted and shows excellent agreement with the calculated depth, consistent with Hertzian contacts.

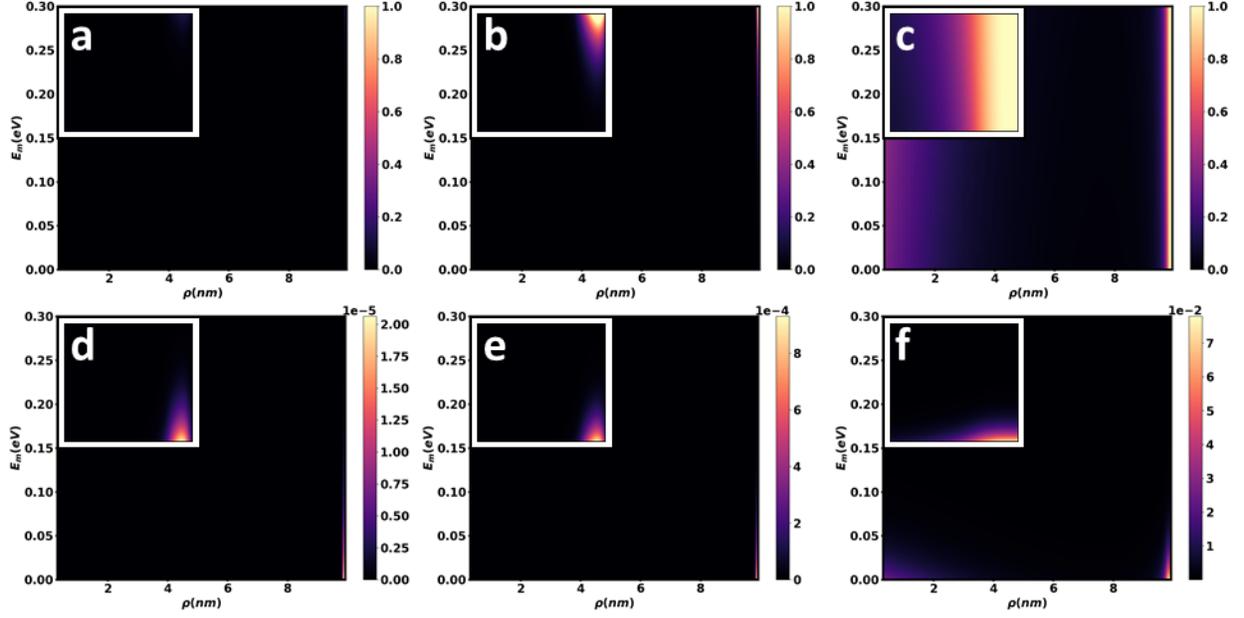

**Fig. S11.**

**a-c**, Tunneling probability $P_{Tun}$ for $F = 3$ µN and $V_{Rev} = 2.19$ V, 3.29 V, and 87.8 V, respectively. The current for the three cases is 0.03 nA, 1 nA, and 300 nA, respectively. These display, in $E_m - \rho$ space, where tunneling dominates. **d-f**, $P_{Tun} \ln\left[\frac{1+\exp(-(E_m+\xi_F)/k_BT)}{1+\exp(-(E_m+\xi_F+qV)/k_BT)}\right]$ for the same reverse biases. These display, in $E_m - \rho$ space, where significant contributions to the current are. The maximum $\rho$ are at the contact radius, $a$. The insets magnify the region from $0.9a \leq \rho \leq a$ and $0 \leq E_m \leq 0.3$ eV. Note that at very low currents, the tunneling only occurs in a small ring-shaped region near the edge of contact.

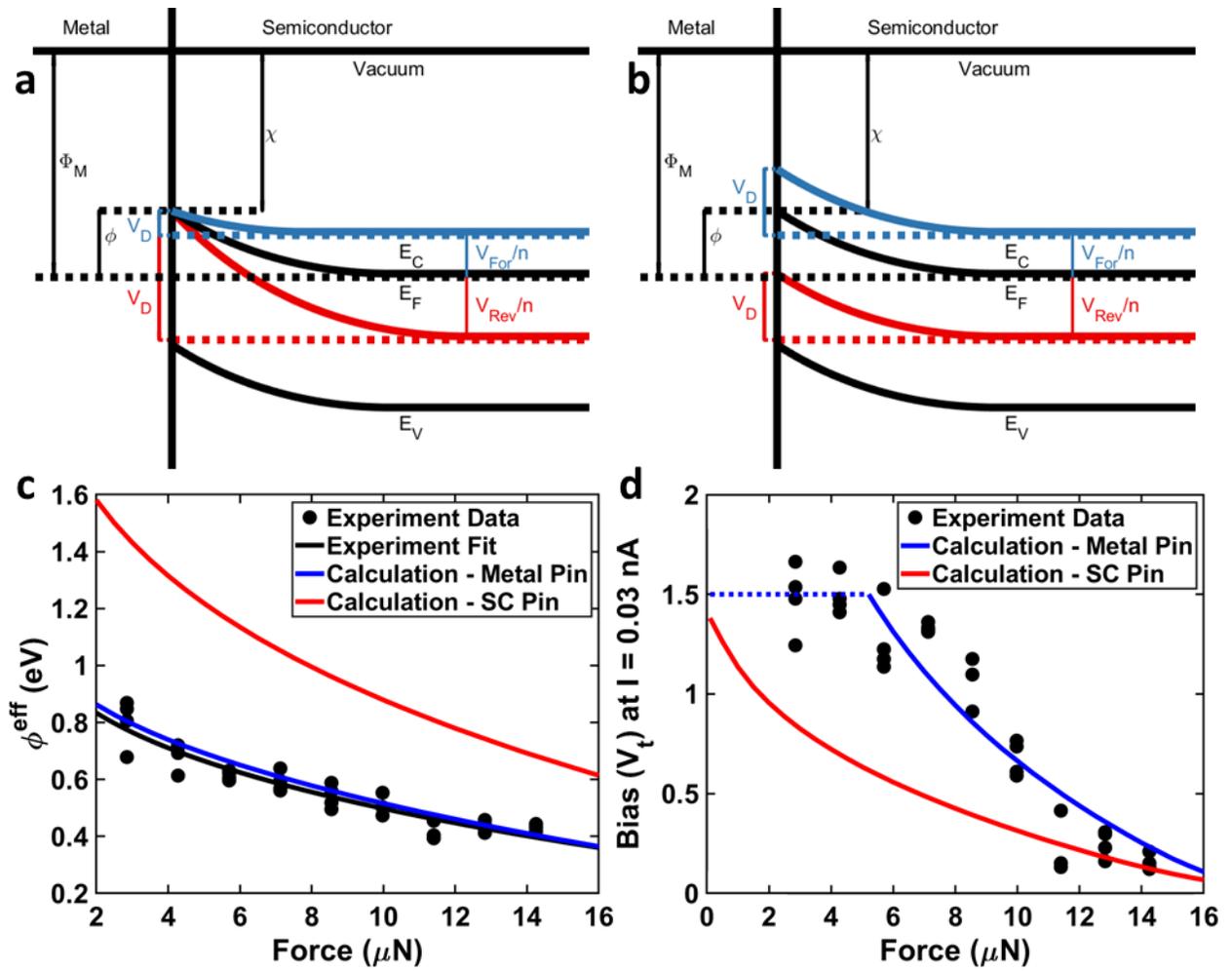

**Fig. S12.**

**a,b**, Band bending diagrams for the case with pinning relative to states in the metal (**a**) (where the pinned states do not shift with applied bias) or the semiconductor (**b**) (where the pinned states shift with applied bias). **c,d**, Calculated $\phi^{\text{eff}}$ (**c**) and reverse bias threshold voltage $V_{\text{Rev,t}}$ (**d**) for pinning in the metal (blue) and semiconductor (red); pinning relative to the semiconductor does not fit the experimental data.

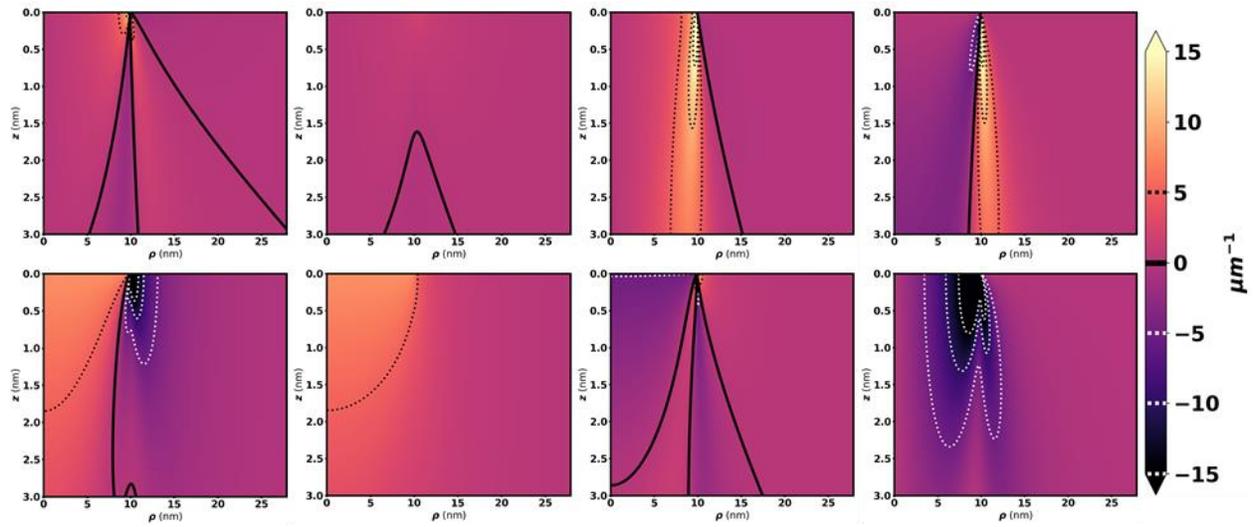

**Fig. S13.**

Strain gradient components for $F = 3$ μN and $R = 60$ nm. Left-to-right, (top) $\varepsilon_{rrr}$, $\varepsilon_{\theta\theta r}$, $\varepsilon_{zzr}$, and $\varepsilon_{rzr}$ and (bottom) $\varepsilon_{rrz}$, $\varepsilon_{\theta\theta z}$, $\varepsilon_{zzz}$, and $\varepsilon_{rzz}$. Contour lines are marked on the color bar.

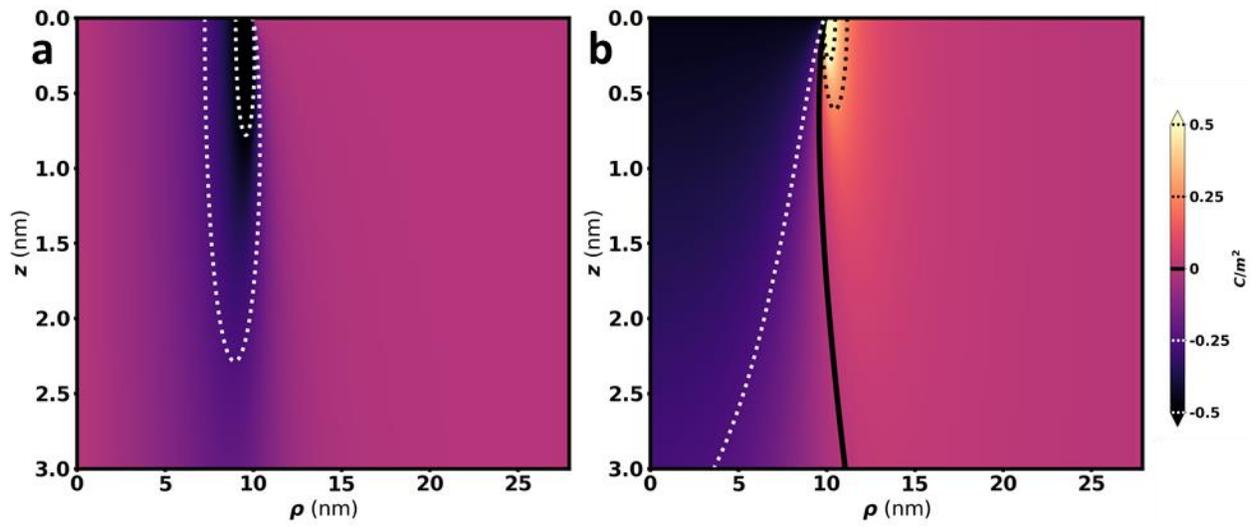

**Fig. S14.**

**a,b**, Polarization components $P_r$ (**a**) and $P_z$ (**b**) for $F = 3\ \mu N$ and $R = 60$ nm.

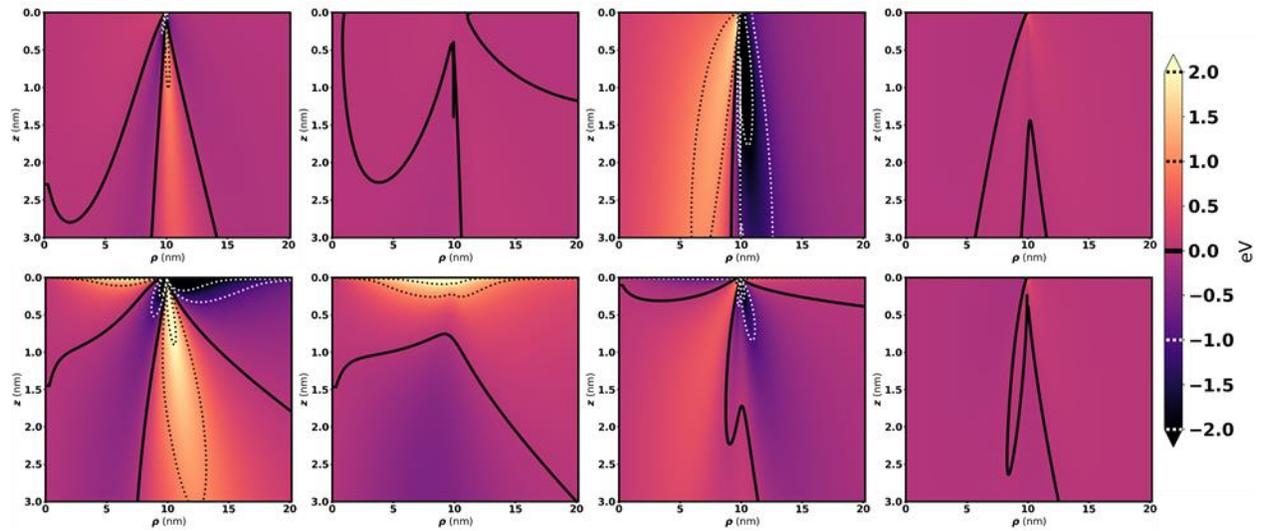

**Fig. S15.**

Contributions of each strain gradient component to the flexoelectric potential $\Phi_{\text{FXE}}$ for $F = 3$ μN and $R = 60$ nm. Left-to-right, contributions from (top) $\varepsilon_{rrr}$, $\varepsilon_{\theta\theta r}$, $\varepsilon_{zzr}$, and $\varepsilon_{rzr}$ and (bottom) $\varepsilon_{rrz}$, $\varepsilon_{\theta\theta z}$, $\varepsilon_{zzz}$, and $\varepsilon_{rzz}$.

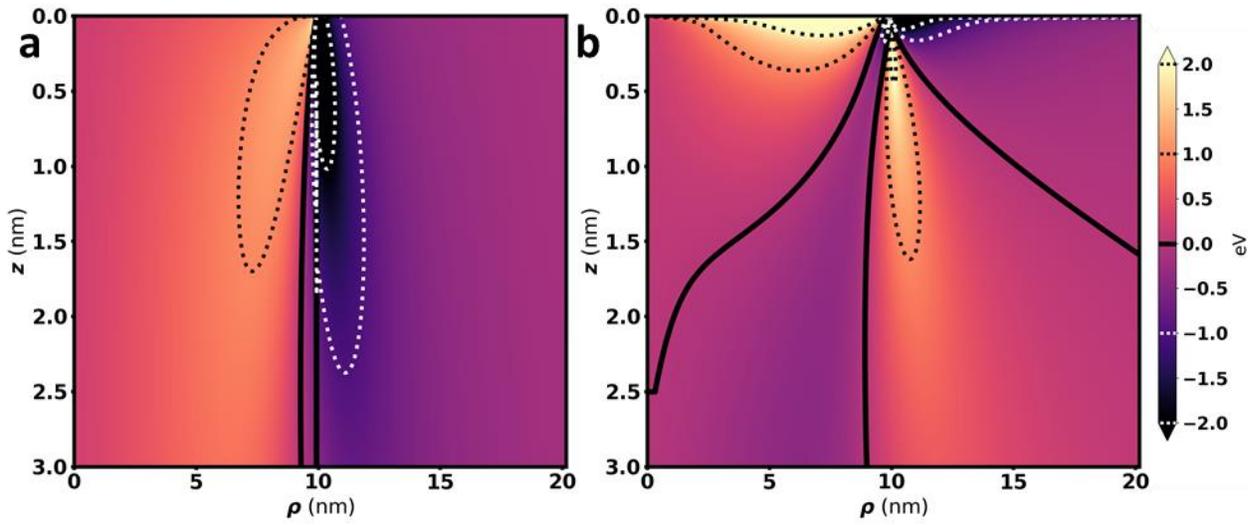

**Fig. S16.**

**a,b**, Contributions to the flexoelectric potential $\Phi_{\text{FXE}}$ for $F = 3$ µN and $R = 60$ nm from the two non-zero polarization directions, $P_r$ (**a**) and $P_z$ (**b**).

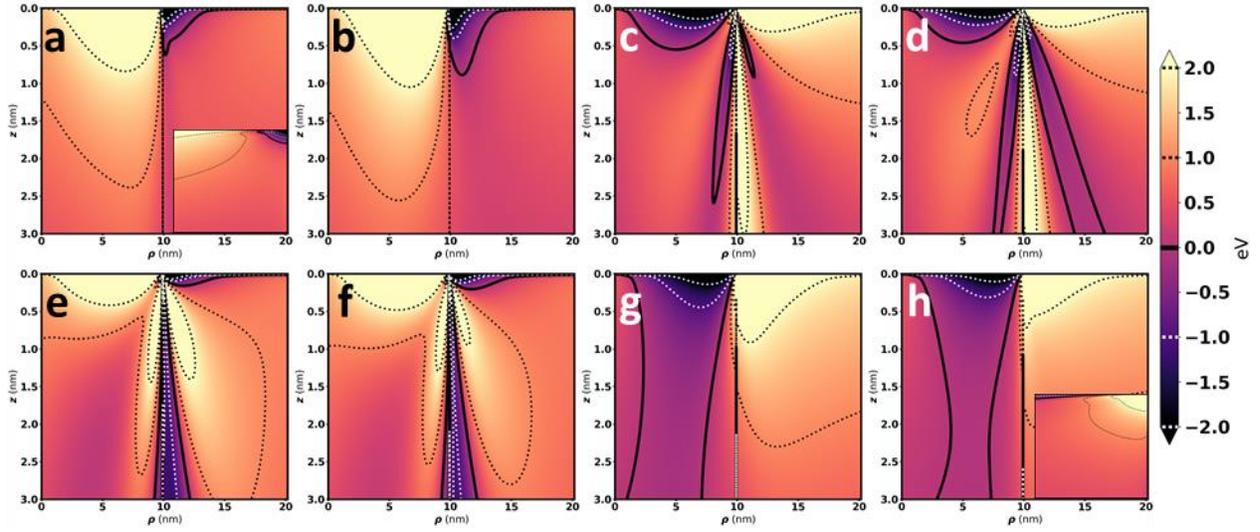

**Fig. S17.**

**a-h**, $E_c(F,r) - E_F$ for $F = 3\ \mu N$ and $R = 60$ nm with the signs of the flexoelectric coefficients artificially modified. The signs are relative to the actual values $\mu_{11} = -36.9$ nC/m, $\mu_{12} = -40.2$ nC/m, and $\mu_{44} = -1.4$ nC/m and the magnitudes stay consistent. **a-d**, $\mu_{11}$, **e-h**, $-\mu_{11}$; **a,b,e,f**, $\mu_{12}$, **c,d,g,h**, $-\mu_{12}$; **a,c,e,g**, $\mu_{44}$, **b,d,f,h**, $-\mu_{44}$. Contour lines are marked on the color bar. **a,h**, (inset) The same at the contact radius ($a$) edge, in the range $0.95a \leq \rho \leq a$.

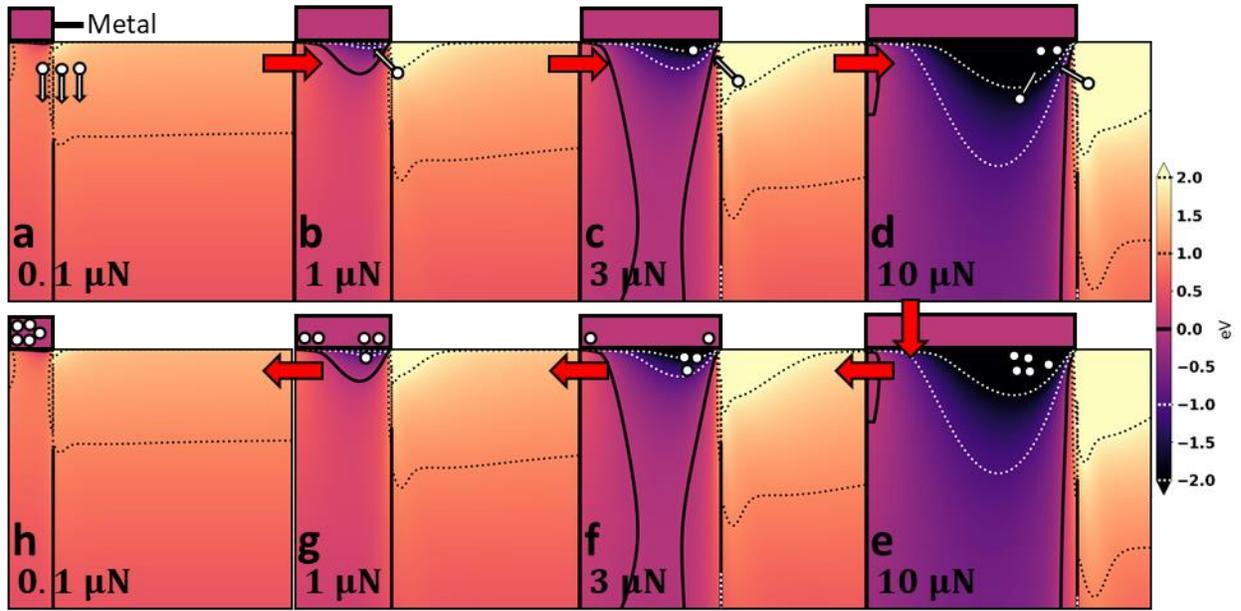

**Fig. S18.**

**a-h**, Ratcheting mechanism for charge transfer in metal-semiconductor contacts. $E_c(F, r) - E_F$ in the range $0 \leq \rho \leq 20$ nm and $0 \leq z \leq 3$ nm with the sign of the flexoelectric coefficients flipped for increasing and decreasing forces $F$, as in contact and pull-off of an asperity. **a**, For very low forces, the depletion potential is most important. **b-d**, As the force increases, electrons (white circles, schematically arrowed) move from regions with an increasing potential to regions with decreasing potential. **e-g**, As the force decreases, the number of available states in the potential well decreases, forcing some electrons into the metal rather than back across the barrier in the semiconductor, where states near the metal Fermi level $E_{F,M} = 0$ are available. **h**, After the force is completely released, electrons have transferred to the metal.

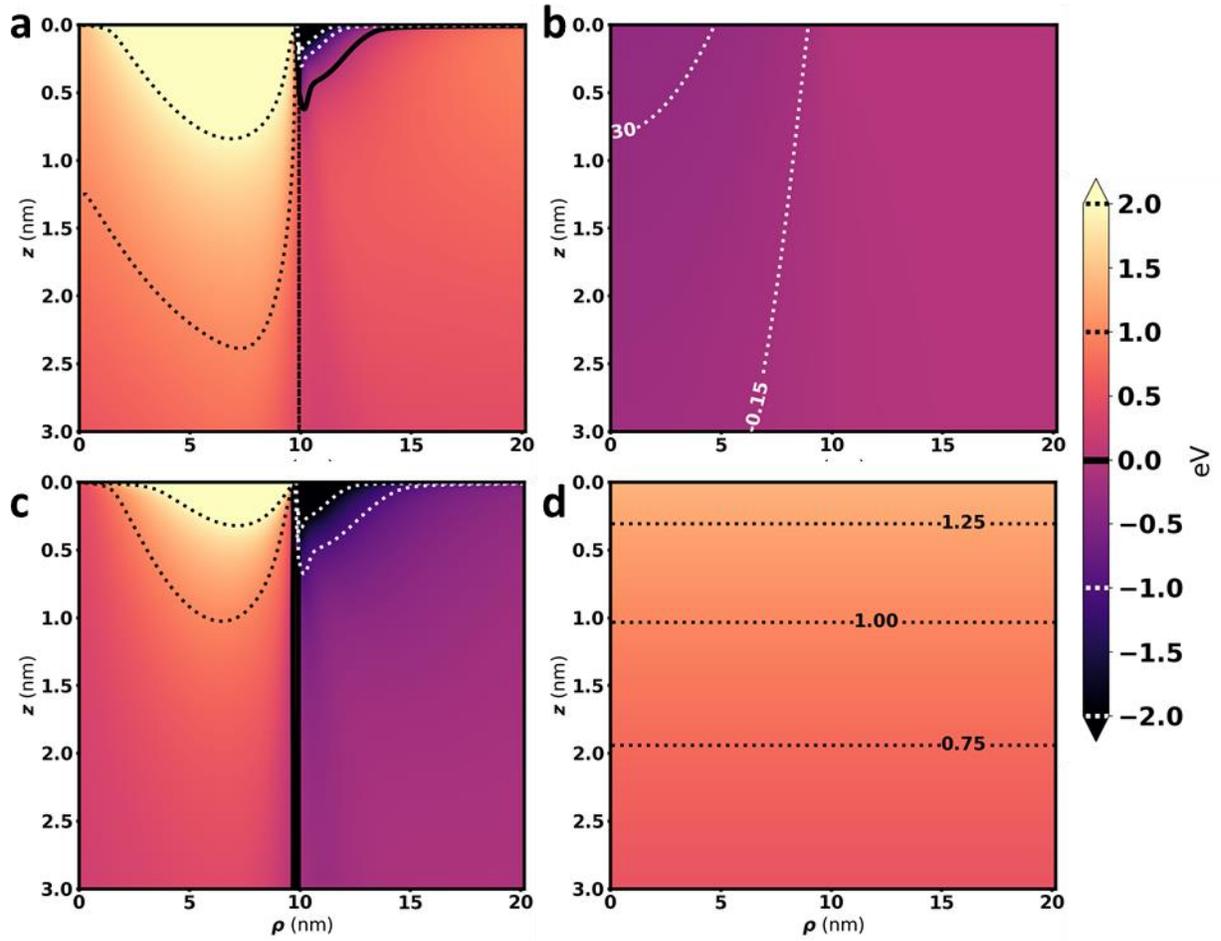

**Fig. S19.**

**a**, $E_c(F, \mathbf{r}) - E_F$ for $F = 3\,\mu\text{N}$ and $R = 60$ nm. **b-d**, Components of $E_c(F, \mathbf{r}) - E_F$: **b**, $\frac{dE_c(F,\mathbf{r})}{d\varepsilon}\varepsilon(F, \mathbf{r})$, **c**, $\Phi_{\text{FXE}}(F, \mathbf{r})$, and **d**, $\Phi_{\text{DEP}}(\mathbf{r})$, from equations (4), (7), and (8), respectively. Unlabeled contour lines are marked on the color bar.

44